\documentclass[%
reprint,
superscriptaddress,
 amsmath,amssymb,
 aps,
]{revtex4-2}

\usepackage{graphicx}
\usepackage{dcolumn}
\usepackage{bm}
\usepackage{xfrac}
\usepackage{comment}
\usepackage{mathtools}
\usepackage{physics}
\usepackage{makecell}
\usepackage{xcolor}

\newcommand{\Ham}{\hat{H}}
\newcommand{\D}{\hat{D}}
\newcommand{\N}{\hat{N}}

\newcommand{\Id}{\hat{I}}
\newcommand{\dm}{\hat{\rho}}
\newcommand{\Miss}{\hat{\mathcal{M}}}

\newcommand{\Lin}{\mathcal{L}}

\begin{document}

\preprint{}

\title{Revealing the physical structure of the general quantum master equation  
}

\author{Eugenia Pyurbeeva}
\email{eugenia.pyurbeeva@mail.huji.ac.il}
\affiliation{The Institute of Chemistry and the Fritz Haber Center for Theoretical Chemistry, The Hebrew University of Jerusalem, Jerusalem 9190401, Israel}
\author{Ronnie Kosloff}
\affiliation{The Institute of Chemistry and the Fritz Haber Center for Theoretical Chemistry, The Hebrew University of Jerusalem, Jerusalem 9190401, Israel}

\date{\today}
\begin{abstract}
The Lindblad (GKLS) master equation, which represents the mathematical form for the general evolution of a density matrix, is a versatile and widely-used tool in open quantum systems. In contrast with the typical approach of imposing additional conditions on the system, such as weak coupling or energy conservation, we explore the structure of the equation with no assumptions. We demonstrate that general quantum dynamics can be expressed through a combination of free evolution, exchanges of some physical quantities (generalised charges), not necessarily commuting with the Hamiltonian, between the system and the bath, and pure dephasing. This result comprises a novel perspective on quantum master equations, employing physical processes as elemental parts. We use it to explore the dynamics and stationary states of a two-level system and show that strong coupling, particle exchange, and non-Abelian effects all share the same physical origin. Moreover, we demonstrate that the generalised Gibbs state for all three cases contains a non-commutation term, which has not been previously considered. 
\end{abstract}
\maketitle
\section{Introduction}
The GKLS master equation is one of the central tools in open quantum systems and quantum thermodynamics. Introduced by Gorini, Kossakowski, Lindblad and Sudarshan \cite{lindblad1976generators,gorini1976completely}, it gives the most general mathematical form of the physically valid (completely positive and trace-preserving) semi-group evolution of a density matrix:
\begin{equation}
  \label{eq:lindblad}
    \frac{\dd \dm}{\dd t}=-i [ \Ham, \dm]+\sum \gamma_i \Big(L_i \dm L_i^\dagger-\frac{1}{2}\{L_i^\dagger L_i, \dm\} \Big)  
\end{equation}
where the rate coefficients $\gamma_{i}$ are real and positive, and $L_i$, known as the Lindblad jump operators are not determined by the framework and can be arbitrary.

The mathematical generality makes the equation extremely versatile -- through choosing $\Ham$, $\gamma_i$'s and  $L_i$ it allows to describe any dynamics, extending even to non-Markovian \cite{dann2022non}. It has been employed in fields as diverse as diverse as quantum optics \cite{allen2012optical}, quantum information processing \cite{rivas2012open} and quantum computing \cite{kashyap2025accuracy}.

However, the generality comes at a cost in connection with physical reality. The equation is highly-degenerate, with a set of non-trivial transforms between $\Ham$, $\gamma_i$'s and  $L_i$ allowed to describe the same dynamics (\cite{breuer2002theory, Manzano2020}, see Appendix \ref{app-transforms}). These include shifting of terms between the free evolution and the dissipative parts, which makes the equation difficult for intuitive interpretation. More importantly, not all choices of $\gamma_i$'s and  $L_i$ lead to physical dynamics -- while the equation always has a fixed point \cite{frigerio1978stationary}, not all choices lead to it being a thermal state. 

The tenuous connection between the GKLS equation and physical reality means that writing it for a known physical system is a non-trivial problem. It is most often approached by imposing additional assumptions on the system, such as the weak coupling limit \cite{davies1974markovian} or strict energy conservation \cite{dann2021open}, which set limitations on the free parameter space. Despite this, explicit expressions of the GKLS equations corresponding to a physical system are few, far-between, and limited to relatively special cases \cite{Smirne2010, kbsf-x4bz}. 
In this work, we take a different approach. Instead of starting from the assumptions imposed on the system intended for the description by the GKLS equation, we study the structure of the equation, preserving its generality. The main result of the paper is a demonstration that arbitrary Lindblad dynamics of a two-level system can be expressed as a sum of free evolution, one or more exchanges of a physical quantity (generalised charge) described by a Hermitian operator with a bath(s), and additional pure dephasing, due to external noise \cite{Kosloff2019} or weak measurement \cite{Korotkov2001}. 

This decomposition gives an immediate physical interpretation to the dynamics. As it is derived from the initial form of the GKLS equation, it is fully general, and valid for systems in regimes typically less amenable to theoretical description, such as those with strong coupling \cite{Anto-Sztrikacs2023, Burke2024} or far from equilibrium \cite{Schaller2014}.  Also, due to the explicit inclusion of exchanges in the equation, it provides a natural framework for the description of a system exchanging quantities with other than energy with the bath, such as describing particle exchange and grand-canonical systems \cite{Site2024, Carlen2025, Reible2025}, as well as the exchange of multiple non-commuting quantities, known as non-Abelian thermodynamics \cite{Guryanova2016, YungerHalpern2016, Lostaglio2017}. 

The paper is structured as follows: in Section \ref{sec-maths} we state our criterion for a ``physical'' exchange, and then proceed to prove the main mathematical result of the work for the case of a two-level system and two Lindblad jump operators. In Section \ref{sec-cases} we explore the resulting dynamics and stationary states for several simple illustrative cases, showing the existence of exceptional points of the dynamics, as well as additional exponential terms in the stationary states, deviating from the expected standard Gibbs states. In Section \ref{sec-physics} we discuss our findings in context of the current results in open quantum systems and potential experimental studies, as well as the extension of our results to arbitrary many-level systems. Finally, in Conclusions, we summarise and outline our vision for the follow-up work.

\section{Mathematical construction}
\label{sec-maths}
\subsection{Physical exchange}
\label{sec:physical-exchange}
The field of ``open quantum systems'' concerns systems open to interactions with their environment. Most commonly, these interactions are understood as an exchange of some physical quantity, a generalised charge. It is often considered to be energy, but can be real or quasi-particles, spin, etc. The aim of this work is to cast the general evolution of the density matrix, as described by the GKLS equation, in terms of such exchanges. 

To this end, we need to give a mathematical definition for the intuitive notion of ``physical exchange''. In our previous work, we showed that for a two-level system under weak coupling and strict energy conservation constraints coupled to a single bath, the Lindblad two jump operators describing the dissipation, $L_p$ and $L_m$, can be rescaled to $\sigma_p$ and $\sigma_m=\sigma_p^\dagger$, obeying the algebraic relations of fermion creation and annihilation operators:
\begin{equation}
\label{eq:fermi-algebra}
    \begin{split}
        \sigma_p^2=\sigma_m^2=0\\
        \{\sigma_p,  \sigma_m\}=\Id
    \end{split}
\end{equation}
From the physical perspective, the system is exchanging energy with a bath, and, due to weak coupling and energy conservation, the energy quanta exchanged were precisely on resonance — equal to the energy level spacing.

The commutator of the jump operators in that case is equal to:
\begin{equation}
[\sigma_p,  \sigma_m]=\frac{\Ham}{E}
\end{equation}
where $\Ham$ is the Hamiltonian of the system and $E$ — the energy level spacing.

By analogy with the weakly-couple energy exchange case outlined above, we arrive at the central idea of the work. 

We consider a Lindblad dissipator to be describing a physical exchange if it is based on a pair of Lindblad jump operators following a fermionic algebra \ref{eq:fermi-algebra}. The generalised charge $\N$, the physical quantity corresponding to the exchange, is then described by the Hermitian operator: 
\begin{equation}
\N=[\sigma_p,  \sigma_m]
\end{equation}
It may seem counter-intuitive for operators following a fermion algebra to describe the exchange of often bosonic quantities, such as energy, but the reason for it lies in the definition of energy levels. In the absence of energy level degeneracy, a single level can be either occupied or empty, with double occupation impossible. Thus, even if the exchanged quantity has boson statistics in the bath, the jump operators for the exchange between any two levels have to be fermionic.

Now, having established a mathematical framework for the exchange of a physical quantity, we proceed to rewrite the GKLS equation in terms of such exchanges. 

\subsection{The equation}
\label{sec:equation}
We start with a general Lindblad equation, for simplicity for a two-level system:
\begin{equation}
\label{eq:Lindblad-start}
    \frac{\dd \rho}{\dd t}=-i [ \Ham, \dm]+\sum_{i=1,2} \gamma_i \left(L_i \rho L_i^\dagger-\frac{1}{2}\{L_i^\dagger L_i, \dm\} \right)
\end{equation}
(higher dimension systems can be decomposed into pairwise dynamics \cite{Pyurbeeva2026}, we will discuss the extensions below).

We consider the case of two Lindblad operators as a single one either describes pure dephasing, if it is Hermitian \cite{dann2021open}, or, if not, the dynamics are unphysical, similar to a single lowering jump operator $\sigma_m$, without a corresponding $\sigma_p$. Such a system still has a fixed point, however it describes a universe in which the reverse process has a zero probability, breaking detailed balance. Thus, two Lindblad operators give the simplest case of non-trivial physical dynamics.

For the same reason as above, we impose a single constraint on the Lindblad jump operators $L_1$ and $L_2$ -- their linear space is closed under the adjoint operation. Or, in other words, $L_1^\dagger$ and $L_2^\dagger$ belong to their linear space.

We can always choose to further restrict ourselves to the space of traceless operators, as the transform $L_i\rightarrow L_i-\alpha_i \Id$, $\Ham \rightarrow \Ham +\sum_i\frac{\gamma_i}{2i} \left(\alpha_i^*L_i-\alpha_iL_i^\dagger \right)$ preserves the dynamics (see Appendix \ref{app-transforms}). 

This allows to us define a linear space of Hermitian operators based on the traceless $L_1+L_1^\dagger$ and $L_2+L_2^\dagger$. If $L_1$ and $L_2$ are non-collinear, the space is two-dimensional, and we can use it to define an in-plane orthonormal basis $A_1$, $A_2$, where $A_1^2=A_2^2=\Id$, $\Tr (A_1A_2)=0$. (Note that $M^2\sim \Id$ is a general property of traceless operators, even non-Hermitian). The third operator to complete the basis, and extend it to the entire three-dimensional space of traceless 2$\times$2 Hermitian operators is $A_3=\frac{1}{2i} [A_1, A_2]$. The choice of $A_3$ is unique (up to the sign), while $A_1$ and $A_2$ have a rotational freedom in the plane. See Appendix \ref{ap:trivial} for traceless operator properties and normalisation of $A_3$. 

In this basis, we write the initial Lindblad operators $L_1$ and $L_2$ as $L_i=a_i A_1 + b_i A_2$, where $a, b \in \mathbb{C}$, and expand the dissipators of the equation (Eq.\ref{eq:Lindblad-start}) as:
\begin{multline}
    L\rho L^\dagger-\frac{1}{2}\{L^\dagger L, \dm \}=\\ \|a \|^2 \Big(A_1 \dm A_1 -\frac{1}{2}\{A_1^2, \dm\} \Big)+\|b \|^2 \Big(A_2 \dm A_2 -\frac{1}{2}\{A_1^2, \dm\} \Big)\\+\text{Re}(a^*b)\Big(A_1\dm A_2 +A_2 \dm A_2- \frac{1}{2}\{A_2 A_1+ A_1 A_1,\dm \}\Big)\\+i \text{Im}(a^*b)\Big(A_1\dm A_2 -A_2 \dm A_2- \frac{1}{2}\{A_2 A_1- A_1 A_2, \dm \}\Big)
\end{multline}
where, for brevity, we have removed the index $i$ enumerating the dissipator terms.

Thus, the complete Lindblad dissipator, the sum of both terms, can be written as:
\begin{multline}
\label{eq:Lindblad-terms}
    \Lin (\dm)=a \Big(A_1 \dm A_1 -\frac{1}{2}\{A_1^2, \dm\} \Big)+b \Big(A_2 \dm A_2 -\frac{1}{2}\{A_1^2, \dm\} \Big)+\\+c\Big(A_1\dm A_2 +A_2 \dm A_1- \frac{1}{2}\{A_2 A_1+ A_1 A_2,\dm \}\Big)+\\+i d\Big(A_1\dm A_2 -A_2 \dm A_1- \frac{1}{2}\{A_2 A_1- A_1 A_2, \dm \}\Big)
\end{multline}
where $a,b,c,d \in \mathbb{R}$. 

\subsubsection{The imaginary term of the dissipator}
First, we look at the imaginary part of the dissipator:
\begin{equation}
    \Lin_{\text{Im}} (\dm)=i \Big(A_1\dm A_2 -A_2 \dm A_2- \frac{1}{2}\{A_2 A_1- A_1 A_2, \dm \}\Big)
\end{equation}

Another property of the traceless operator space is that the anticommutator is proportional to identity (see Appendix \ref{ap:trivial}). Applying it to the commutators of $A_1$ and $A_2$ with the density matrix, we obtain:
\begin{equation}
    \begin{split}
        \{A_1, \dm \}=\{A_1, \frac{\Id}{2}+\dm_0 \}=A_1+\alpha_1 \Id\\
        \{A_2, \dm \}=\{A_2, \frac{\Id}{2}+\dm_0 \}=A_2+\alpha_2 \Id
    \end{split}
\end{equation}
where $\dm_0$ is the reduced traceless density matrix, and $\alpha_1, \alpha_2$ -- the complex numerical coefficients. The anticommutator relations allow us to swap the multiplicative terms and expand:
\begin{multline}
    A_1 \dm A_2=\\=\big(A_1-\alpha_1 \Id-\dm A_1 \big)A_2=A_1 A_2+\alpha_1 A_2-\dm A_1 A_2=\\=A_1\big(A_2-\alpha_2 \Id-\dm A_2 \big)=A_1 A_2+\alpha_2 A_1- A_1A_2\dm=\\=
    A_1A_2+\frac{1}{2}\big(\alpha_2 A_1+ \alpha_1 A_2\big)-\frac{1}{2}\big(\dm A_1 A_2 +A_1 A_2 \dm\big)
\end{multline}
Using the same procedure:
\begin{multline}
    A_2 \dm A_1=\\A_2 A_1+\frac{1}{2}\big(\alpha_1 A_2+ \alpha_2 A_1\big)-\frac{1}{2}\big(\dm A_2 A_1 +A_2 A_1 \dm\big)
\end{multline}
Taking the difference between the terms, we arrive at:
\begin{multline}
\label{eq:cross}
    A_1 \dm A_2-A_2 \dm A_1=A_1A_2-A_2A_1+\\+\frac{1}{2}\big((A_2A_1-A_1A_2)\dm+\dm(A_2A_1-A_1A_2) \big)=\\=[A_1,A_2]-\frac{1}{2}\{[A_1, A_2], \dm \}
\end{multline}
Combined with the $-\frac{1}{2}\{A_2 A_1- A_1 A_2, \dm \}$ term in the imaginary part of the dissipator (note the change of order in the commutator), this leads to dramatic simplification:
\begin{equation}
    \Lin_{\text{im}} (\dm)=i[A_1, A_2]=-2A_3
\end{equation}
\subsubsection{The real term of the dissipator}
The real part of the dissipator can be written as:
\begin{equation}
\label{eq:dis-real}
    \Lin_{\text{R}}=\sum_{i,j}M_{ij}\Big(A_i \dm A_j-\frac{1}{2}\{A_j A_i, \dm \} \Big)
\end{equation}
where $M$ is a real-valued symmetric matrix (see Eq. \ref{eq:Lindblad-terms}). 
As such, it can be diagonalised by an in-plane rotation, meaning that the original Lindblad equation, including both the real and imaginary terms can be written as:
\begin{equation}
    \label{eq:Lindblad-A0}
    \frac{\dd \rho}{\dd t}=-i [ \Ham, \dm]+\sum_{i=1,2}\gamma_i\Big(A_i \dm A_i -\frac{1}{2}\{A_i^2, \dm \} \Big)+\gamma_3 A_3
\end{equation}
where the in-plane basis $A_1, A_2$ has been rotated, and the imaginary part depends only on $A_3$, which is not affected by the rotation.

The equation \ref{eq:Lindblad-A0} has direct physical interpretation. The ``noise'' terms $A_i \dm A_i -\frac{1}{2}\{A_i^2, \dm \}$ represent dephasing orthogonal to $A_1$ and $A_2$ at differing rates \cite{Kosloff2019, dann2021open}, while the ``force'' term $A_3$ drives the density matrix in its direction. The system will reach a stationary state when the constant drive towards $A_3$ will be compensated by reduction of the $A_3$ component of the density matrix due to the dephasing terms.  

Before we proceed, we address another question. The matrix describing the full Equation \ref{eq:Lindblad-terms} with the real and imaginary parts in the form of Eq. \ref{eq:dis-real} is Hermitian, and thus has two real eigenvalues and can be diagonalised with a unitary transform preserving Lindblad dynamics (see Appendix \ref{app-transforms}). Why did we not do it? The reason for it is that if the imaginary part is non-zero, the unitary transformation required to diagonalise the matrix will have complex-valued coefficients, and the thus eigenvectors will no longer be Hermitian operators. Separating the dynamics into the real and imaginary parts allowed us to remain in the Hermitian basis. 


\subsection{The exchange dissipators}
\subsubsection{The choice of jump operators}
In the previous section (Section \ref{sec:equation}) we found a form of the GKLS equation for a two-level system with two jump operators in an orthonormal basis $\{A_1, A_2, A_3 \}$ (Eq. \ref{eq:Lindblad-A0}). Our aim is to express it 
through an exchange process with jump operators $\sigma_p$ and $\sigma_m$ following fermionic relations, and to this end we find $\sigma_p$ and $\sigma_m$ in the same basis. We begin with a guess:
\begin{equation}
\label{eq:sigmas}
    \begin{dcases}
        \sigma_p=\frac{1}{2}\left(A_1+i A_2\right)\\
        \sigma_m=\frac{1}{2}\left(A_1-i A_2\right)
    \end{dcases}
\end{equation}
and show that $\sigma_p$ and $\sigma_m$ in this form satisfy the requirements for a fermionic algebra (Eq. \ref{eq:fermi-algebra}):

First, trivially, $\sigma_p=\sigma_m^\dagger$. As $\sigma_p$ (and $\sigma_m$) is traceless, $\sigma_p^2\sim \Id$. Additionally, 
    \begin{multline}
        \Tr \sigma_p^2=\frac{1}{4}\Tr((A_1+i A_2)(A_1+i A_2))=\\=\frac{1}{4}(\Tr A_1^2+2i\Tr(A_1A_2)-\Tr A_2^2)=0
    \end{multline}
as $A_1$ and $A_2$ are normalised and $\Tr(A_1A_2)=0$. This proves that $\sigma_p^2=\sigma_m^2=0$.

Moreover,
\begin{multline}
    \{\sigma_p, \sigma_m \}=\frac{1}{4}\{(A_1+i A_2),(A_1-i A_2)\}=\\=\frac{1}{4}\left( \{A_1, A_1\}+\{A_2, A_2\}+2i\{A_1, A_2\}\right)=\\=\frac{1}{4}\left(2A_1^2+2A_2^2\right)=1
\end{multline}

And finally,
\begin{multline}
    [\sigma_p, \sigma_m]=\frac{1}{4}[(A_1+i A_2),(A_1-i A_2)]=\\=\frac{1}{4}\left(i[A_2, A_1]-i[A_1, A_2] \right)=-2i[A_1, A_2]=A_3  
\end{multline}
We also note that our choice of $\sigma_p$ and $\sigma_m$ is unique up to trivial transformations of swaps, multiplying by $e^{i\phi}$, and rotations. We consider the former two irrelevant and keep the latter freedom in mind to be addressed below. 

\subsubsection{The dissipators}
Using the expression of the fermionic jump operators $\sigma_p$ and $\sigma_m$ through the Hamiltonian basis (Eq.\ref{eq:sigmas}), we can expand the exchange dissipators from their standard form:
\begin{equation}
\label{eq:diss-standard}
    \begin{dcases}
    \Lin_p(\dm)=\sigma_p \dm \sigma_m-\frac{1}{2}\{\sigma_m\sigma_p, \dm\}\\
    \Lin_m(\dm)=\sigma_m \dm \sigma_p-\frac{1}{2}\{\sigma_p\sigma_m, \dm\}
    \end{dcases}
\end{equation}
\\
into the basis Hermitian operators:
\begin{multline}
    \Lin_p(\dm)=\frac{1}{4}\left(A_1\dm A_1 + A_2\dm A_2\right) -\frac{\dm}{2}-\\-\frac{i}{4}\left(A_1\dm A_2 - A_2\dm A_1 \right)+\frac{1}{4}\{A_3, \dm \}
\end{multline}
\begin{multline}
    \Lin_m(\dm)=\frac{1}{4}\left(A_1\dm A_1 + A_2\dm A_2\right) -\frac{\dm}{2}+\\+\frac{i}{4}\left(A_1\dm A_2 - A_2\dm A_1 \right)-\frac{1}{4}\{A_3, \dm \}
\end{multline}
Using the previous result for the cross-term $A_1\dm A_2 - A_2\dm A_1$ (Eq.\ref{eq:cross}), this can be further simplified to:
\begin{equation}
\label{eq:diss-new}
        \Lin_{p/m}(\dm)=\frac{1}{4}\left(A_1\dm A_1 + A_2\dm A_2\right)-\frac{\dm}{2}\pm \frac{A_3}{2}
\end{equation}

\subsection{Comparing to previous results}
In a previous paper \cite{Pyurbeeva2026}, we have shown  that under the conditions of weak coupling and strict energy conservation, which requires $\sigma_p$ and $\sigma_m$ to be eigenoperators of the commutator with the Hamiltonian $[\Ham, \bullet]$, the plus and minus parts of the dissipator can be written as:
\begin{equation}
\label{eq:diss-old}
    \Lin^E_{p/m}=-(\rho-\frac{1}{2}\Id)\pm \frac{\Ham}{E}+\frac{[\Ham, [\Ham, \dm]]}{2E^2}
\end{equation}
In that case, we had imposed a scaling of $[\sigma_p, \sigma_m]=2\Ham/E$, and physically, the weakly-coupled system
was exchanging quanta of energy exactly on resonance with the energy level splitting of the system. For a more general exchange, as discussed above (Sec. \ref{sec:physical-exchange}), we believe that the commutator of the fermionic jump operators represents the physical quantity exchanged between the bath and the system.  

Inserting the current commutator value $[\sigma_p, \sigma_m]=A_3$ into Equation \ref{eq:diss-old}, we obtain:
\begin{equation}
\label{eq:diss-A3}
    \Lin^E_{p/m}=-(\rho-\frac{1}{2}\Id)\pm\frac{A_3}{2}+\frac{[A_3, [A_3, \dm]]}{8}
\end{equation}
We show that this is equivalent to the current result (Eq.\ref{eq:diss-new}).

The $\pm A_3/2$ terms obviously agree. Expanding the double dissipator term for an arbitrary (Hermitian and normalised) operator $M$, we get:
\begin{multline}
    \frac{[M,[M,\dm]]}{2}=\\=\frac{1}{2}\left(M^2 \dm -2M\dm M +\dm M^2 \right)=\dm - M\dm M
\end{multline}
This shows that the $A_i\dm A_i-\dm$ terms in Eq.\ref{eq:Lindblad-A0} represent dephasing that diminishes all components of the density matrix orthogonal to $A_i$ \cite{Kosloff2019, dann2021open}. 

The dissipator in Eq.\ref{eq:diss-A3} is expressed through $A_3$ only, while our result (Eq.\ref{eq:diss-new}) has terms involving $A_1$ and $A_2$. To proceed further, we find a relation between the ``sandwich'' terms: $A_1\dm A_1$, $A_2\dm A_2$ and $A_3\dm A_3$. As $A_1$, $A_2$, and $A_3$ are a basis in the traceless Hermitian operator space, the density matrix $\dm$ can be represented as: $\dm=\Id/2+r_1 A_1 + r_2 A_2+ r_3 A_3$. Then, since the basis vectors follow the Pauli matrix algebra, we have:
\begin{equation}
    \begin{dcases}
        A_1\dm A_1=\frac{1}{2}\Id+r_1 A_1-r_2 A_2 - r_3 A_3\\
        A_2\dm A_2=\frac{1}{2}\Id-r_1 A_1+r_2 A_2 - r_3 A_3\\
        A_3\dm A_3=\frac{1}{2}\Id-r_1 A_1-r_2 A_2 + r_3 A_3
    \end{dcases}
\end{equation}
Adding all the terms up, we obtain:
\begin{equation}
    A_1\dm A_1+A_2\dm A_2+A_3\dm A_3=2\Id-\dm
\end{equation}
This allows to express the double commutator in Eq.\ref{eq:diss-A3} as:
\begin{multline}
    \frac{[A_3, [A_3, \dm]]}{8}=\frac{1}{4}\left(\dm-A_3 \dm A_3 \right)=\\=\frac{1}{4}\left(\dm+A_1\dm A_1+A_2\dm A_2 -2\Id +\dm \right)=\\=\frac{\dm}{2}-\frac{1}{2}\Id+\frac{A_1\dm A_1+A_2\dm A_2}{4}
\end{multline}
This shows the equivalence between our current and previous results. 

\subsection{The general equation representation}
As $A_i^2=\Id$, the general form of the Lindblad equation can be simplified to:
\begin{equation}
\label{eq:Lindblad-final}
    \frac{\dd \rho}{\dd t}=-i [ \Ham, \dm]+\gamma_1\left(A_1 \dm A_1-\dm \right)+\gamma_2\left(A_2 \dm A_2-\dm \right)+\gamma_3 A_3
\end{equation}
Our aim was to represent it as an exchange process:
\begin{equation}
    \frac{\dd \rho}{\dd t}=-i [ \Ham, \dm]+\gamma_p \Lin_p(\dm)+\gamma_m \Lin_m(\dm)
\end{equation}
where $\Lin_{p/m}$ are standard Lindblad plus and minus dissipator components (Eq.\ref{eq:diss-standard}) with $\sigma_p$ and $\sigma_m$ following fermionic relations (Eq.\ref{eq:fermi-algebra}). We have found that in our choice of orthonormal basis:
\begin{equation}
        \Lin_{p/m}(\dm)=\frac{1}{4}\left(A_1\dm A_1 + A_2\dm A_2\right)-\frac{\dm}{2}\pm \frac{A_3}{2}
\end{equation}
Now we express the GKLS equation (Eq. \ref{eq:Lindblad-final}) through $\Lin_{p/m}$. We can always choose $\gamma_p-\gamma_m=2 \gamma_3$ to equalise the driving term, however the rest seems impossible. Any combination of $\Lin_p$ and $\Lin_m$ will be symmetric in $A_1 \dm A_1$ and  $A_2 \dm A_2$, while the general evolution (Eq.\ref{eq:Lindblad-final}) is not. This limitation is physical, as thermal dephasing in an exchange process is always symmetric.  

However, we can choose $\gamma_p+\gamma_m=\text{min}(\gamma_1, \gamma_2)$, and have the remaining ``noise'' term as additional dephasing. This gives us the final, fully general result for the Lindblad equation as:
\begin{equation}
    \frac{\dd \rho}{\dd t}=-i [ \Ham, \dm]+\gamma_p \Lin_p(\dm)+\gamma_m \Lin_m(\dm)-\Gamma\frac{[\D, [\D, \dm]]}{2}
\end{equation}

Or, using the result for the exchange dissipator from our previous work \cite{Pyurbeeva2026} (Eq. \ref{eq:diss-old}):
\begin{multline}
\label{eq:exchange}
    \frac{\dd \rho}{\dd t}=-i [ \Ham, \dm]-(\gamma_p+\gamma_m)(\rho-\frac{1}{2}\Id)+(\gamma_p-\gamma_m)\N+\\+(\gamma_p+\gamma_m)\frac{[\N, [\N, \dm]]}{2}-\Gamma \frac{[\D, [\D, \dm]]}{2}
\end{multline}

This shows that completely general density dynamics for two Lindblad operators can be uniquely expressed through three operators: $\Ham$, the Hamiltonian, representing free evolution (note that this is the only way to define the Hamiltonian of the system); $\N$, the quantity being exchanged; and $\D$, the additional dephasing operator, orthogonal to $\N$. Finally, we note that in this form, the expression is independent of the choice of rotational orientation of the Hamiltonian basis. 

\section{Case studies}
\label{sec-cases}
The mathematical construction presented in Section \ref{sec-maths} states that any Lindblad evolution described by two jump operators forming a linear space closed to the adjoint can be uniquely expressed as a sum of free evolution with the system Hamiltonian $\Ham$, a physical exchange of a generalised charge, described by an operator $\N$, not necessarily commuting with $\Ham$, between the system and a bath, and additional pure dephasing based on an operator $\D$, orthogonal to $\N$. 

In order to understand the role of these three parameters on the physics of the system, in this section we explore the time evolution dynamics, including exceptional points, separating underdamped from overdamped dynamics \cite{am2015exceptional}, and the stationary states of the density matrix in several simple representative cases. Before we do it, we define the commutation relations between the three operators of interest. 

First, we rewrite our result (Eq. \ref{eq:exchange}) as:
\begin{multline}
    \frac{\dd \rho}{\dd t}=-i E [ \Ham, \dm]-(\gamma_p+\gamma_m)(\rho-\frac{1}{2}\Id)+\\+(\gamma_p-\gamma_m)\N+(\gamma_p+\gamma_m)\frac{[\N, [\N, \dm]]}{2}-\Gamma \frac{[\D, [\D, \dm]]}{2}
\end{multline}
where we have chosen $\Ham$, $\N$ and $\D$ to be dimensionless. From our proof, $\N \perp \D$. We also choose $|\Ham |=|\N|=|\D|=1/2$. The reason for this normalisation choice is two-fold. First, if the full, unnormalised system Hamiltonian is equal to $\Ham_{\text{sys}}=E \Ham$ it gives $E$ as difference between the energy levels of the system. Second, it sets the double commutation relations between $\N$ and $\D$ to:
\begin{equation}
    \begin{split}
        [\D, [\D, \N]]=\N\\
        [\N, [\N, \D]]=\D
    \end{split}
\end{equation}

\subsection{Standard thermalisation}
We start with the case of a two-level system exchanging energy with a bath in the weak coupling limit, where the energy exchange can only occur on resonance. Thus, the generalised charge $\N$ is equal to $\Ham$. Additionally, we consider pure dephasing to be absent ($\D=0$). 

The equation reduces to:
\begin{multline}
\label{eq:case1}
    \frac{\dd \rho}{\dd t}=-i E [ \Ham, \dm]-(\gamma_p+\gamma_m)(\rho-\frac{1}{2}\Id)+\\+(\gamma_p-\gamma_m)\Ham+(\gamma_p+\gamma_m)\frac{[\Ham, [\Ham, \dm]]}{2}
\end{multline}
(identical to \cite{Pyurbeeva2026}).

Using $\N$ (in our case, $\Ham$), $\D$, and $i [\Ham, \D]$ as an orthogonal Hermitian basis, we write the density matrix as:
\begin{equation}
    \dm=\frac{\Id}{2}+\beta \Ham+ \alpha \D + i \lambda [\Ham, \D]
\end{equation}
By component coefficients, the master equation (Eq.\ref{eq:case1}) reduces to:
\begin{equation}
\label{eq:case1-evolution}
    \begin{dcases}
        \frac{\dd \beta}{\dd t}=-(\gamma_p+\gamma_m)\beta+(\gamma_p-\gamma_m)\\
        \frac{\dd \alpha}{\dd t}=\lambda E-\frac{(\gamma_p+\gamma_m)}{2}\alpha\\
        \frac{\dd \lambda}{\dd t}=-\alpha E-\frac{(\gamma_p+\gamma_m)}{2}\lambda 
    \end{dcases}
\end{equation}
This gives the stationary state as: $\alpha=\lambda=0$,\\ $\beta=(\gamma_p-\gamma_m)/(\gamma_p+\gamma_m)$, or
\begin{equation}
    \dm_{\text{st}}=\frac{\Id}{2}+\frac{\gamma_p-\gamma_m}{\gamma_p+\gamma_m}\Ham
\end{equation}
which agrees with the standard Gibbs state \cite{Pyurbeeva2026}. 

Finally, the characteristic matrix of the equation (found from Eq.\ref{eq:case1-evolution}) is:
\begin{equation}
M_0=
    \begin{pmatrix}
    -\gamma & 0 & 0\\[3pt]
    0 & -\frac{\gamma}{2} & E\\[3pt]
    0 & -E & -\frac{\gamma}{2}
\end{pmatrix}
\end{equation}
where $\gamma=\gamma_p+\gamma_m$. The eigenvalues of $M_0$ are: $-\gamma$ and $-\gamma/2\pm i E$, giving one purely decaying and two partially oscillatory modes. Additionally, all eigenvectors are separate and the dynamics have no exceptional points.

\subsection{$\N=\Ham$, additional orthogonal dephasing}
Now, remaining in the weak coupling and strict energy conservation regime, where $\Ham=\N$, we introduce additional dephasing orthogonal to $\N$:
\begin{multline}
\label{eq:case2}
    \frac{\dd \rho}{\dd t}=-i E [ \Ham, \dm]-(\gamma_p+\gamma_m)(\rho-\frac{1}{2}\Id)+(\gamma_p-\gamma_m)\Ham+\\+(\gamma_p+\gamma_m)\frac{[\Ham, [\Ham, \dm]]}{2}-\Gamma \frac{[\D, [\D, \dm]]}{2}
\end{multline}
This gives the density matrix component evolution equations as:
\begin{equation}
\label{eq:case2-evolution}
    \begin{dcases}
        \frac{\dd \beta}{\dd t}=-(\gamma_p+\gamma_m+\frac{\Gamma}{2})\beta+(\gamma_p-\gamma_m)\\
        \frac{\dd \alpha}{\dd t}=\lambda E-\frac{(\gamma_p+\gamma_m)}{2}\alpha\\
        \frac{\dd \lambda}{\dd t}=-\alpha E-\frac{(\gamma_p+\gamma_m+\Gamma)}{2}\lambda 
    \end{dcases}
\end{equation}
The stationary state then is: $\alpha=\lambda=0$, \\ $\beta=(\gamma_p-\gamma_m)/(\gamma_p+\gamma_m+\Gamma/2)$, or
\begin{equation}
    \dm_{\text{st}}=\frac{\Id}{2}+\frac{\gamma_p-\gamma_m}{\gamma_p+\gamma_m+\frac{\Gamma}{2}}\Ham
\end{equation}
Thus the presence of additional orthogonal dephasing changes the effective temperature, while preserving the Gibbs form.

The characteristic matrix for the case is: 
\begin{equation}
M_1=
    \begin{pmatrix}
    -\gamma-\frac{\Gamma}{2} & 0 & 0\\[3pt]
    0 & -\frac{\gamma}{2} & E\\[3pt]
    0 & -E & -\frac{\gamma+\Gamma}{2}
\end{pmatrix}
\end{equation}

The eigenvalues of $M_1$ are 
\begin{equation}
    \begin{dcases}
        \lambda_1=-\gamma-\frac{\Gamma}{2}\\
        \lambda_{2/3}=-\frac{\gamma}{2}-\frac{\Gamma\pm\sqrt{\Gamma^2-16E^2}}{4}
    \end{dcases}
\end{equation}
$\Gamma$, $\gamma$ and $E^2$ are real and above zero, and the sign of $\sqrt{\Gamma^2-16E^2}$ determines the behaviour of the system. If the additional dephasing is weak and $\Gamma<4E$, the system still has one purely decaying mode and two oscillatory modes with complex eigenvalues, as in the case of standard thermalisation, while for strong extra dephasing, $\Gamma>4E$, all three modes describe exponential decay to equilibrium. At the transition point, $\Gamma=4E$, the eigenvalues are equal to: $\lambda_1=-\gamma-\Gamma/4$ and $\lambda_2=\lambda_3-\gamma/2-\Gamma/4$, which follows the 2,1,1 ratio of pure thermalisation at $E=0$.

\subsection{$\Ham\neq\N$, no additional dephasing}
\begin{figure}
\includegraphics[width=\linewidth]{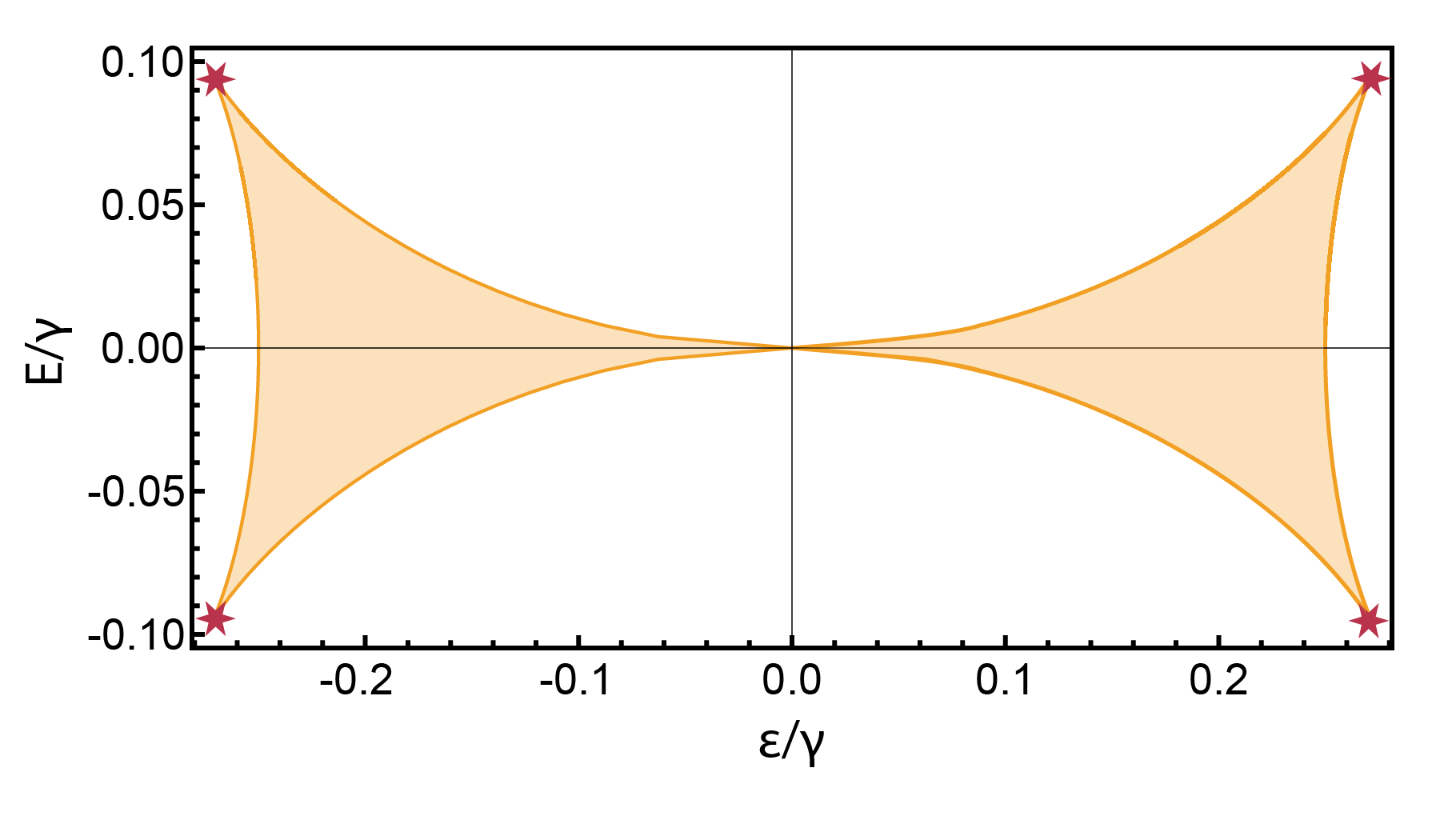}
\caption{A map of the types of dynamics of the system with non-commuting exchange operator, $N$, and Hamiltonian, $\Ham$ (Eq.\ref{eq:case3}). Within the shaded areas all three eigenvalues are real and the dynamics are of pure decay. The lines show second-order exceptional points, separating areas with real and complex eigenvalues, while the cusps (marked with red stars), at $E/\gamma=\pm 1/(6\sqrt{3})$, $\varepsilon/\gamma=\pm\sqrt{2}/(3\sqrt{3})$ represent third-order exceptional points. 
\label{fig-expoints}}
\end{figure}
Finally, we look at the case where the free evolution Hamiltonian differs from the exchange quantity $\N$ and does not commute with it. For simplicity, we assume that it has a small additional contribution in the $\D$ direction, while the pure orthogonal dephasing is absent:
\begin{multline}
\label{eq:case3}
    \frac{\dd \rho}{\dd t}=-i [ E\N+\varepsilon \D, \dm]-(\gamma_p+\gamma_m)(\rho-\frac{1}{2}\Id)+\\+(\gamma_p-\gamma_m)\N+(\gamma_p+\gamma_m)\frac{[\N, [\N, \dm]]}{2}
\end{multline}
(In this case $E$ is no longer the energy level spacing, but we write the master equation in this way for simplicity).

Then the density matrix component evolution equations are:
\begin{equation}
\label{eq:case3-evolution}
    \begin{dcases}
        \frac{\dd \beta}{\dd t}=-(\gamma_p+\gamma_m)\beta-\varepsilon \lambda+(\gamma_p-\gamma_m)\\
        \frac{\dd \alpha}{\dd t}=\lambda E-\frac{(\gamma_p+\gamma_m)}{2}\alpha\\
        \frac{\dd \lambda}{\dd t}=-\alpha E+ \varepsilon \alpha-\frac{(\gamma_p+\gamma_m)}{2}\lambda 
    \end{dcases}
\end{equation}
The stationary values of the coefficients then are:
\begin{equation}
\label{eq:case3-stst}
    \begin{dcases}
        \beta=\frac{\gamma_p-\gamma_m}{\gamma_p+\gamma_m} \frac{4E^2+(\gamma_p+\gamma_m)^2}{(4E^2+(\gamma_p+\gamma_m)^2+2\varepsilon^2)}\\
        \alpha=\frac{\gamma_p-\gamma_m}{\gamma_p+\gamma_m} \frac{4\varepsilon E}{(4E^2+(\gamma_p+\gamma_m)^2+2\varepsilon^2)}\\
        \lambda=(\gamma_p-\gamma_m) \frac{4\varepsilon}{(4E^2+(\gamma_p+\gamma_m)^2+2\varepsilon^2)}
    \end{dcases}
\end{equation}
meaning that the stationary state has all three vector components. Note that if $\varepsilon=0$, it returns to the expected Gibbs state.

Finally, we look at the characteristic matrix:
\begin{equation}
M_2=
    \begin{pmatrix}
    -\gamma & 0 & -\varepsilon\\[3pt]
    0 & -\frac{\gamma}{2} & E\\[3pt]
    \varepsilon & -E & -\frac{\gamma}{2}
\end{pmatrix}
\end{equation}
The dynamics of the system have a non-trivial set of second- and third-order exceptional points, similarly to  \cite{am2015exceptional}, shown in Fig.\ref{fig-expoints}. See Appendix \ref{ap-exp} for the identification of exceptional points of the dynamics.

\section{Discussion: Physical implications}
\label{sec-physics}
\subsection{Master equation structure}
From a mathematical perspective, the main result of this work, Eq.\ref{eq:exchange} is limited to a two-level system described by a GKLS equation with two Lindblad jump operators. In the proof, we also imposed a condition that their linear space is closed under the adjoint operation. For the sake of conciseness, we leave the extension to more complex cases, including more Lindblad operators (the main question there is the uniqueness of the representation) and 
higher-dimensional systems with many energy levels, which can be done by separating the dynamics into pairwise projections, similarly to \cite{Pyurbeeva2026}, for further mathematical work.

However, our results in their current form allow us to make a more general claim about the structure of quantum master equations. We have shown that in our simple case the interaction of a quantum system with the environment is always represented as a combination of exchange of a generalised charge with a bath and ``noise'', or pure dephasing. This can be extended to a more general case. For a more complex many-level system the dynamics may involve multiple exchanges, which may or may not be correlated between the levels, but the structure of a combination of free evolution, exchanges, and pure dephasings will remain. 

Moreover, we argue that this description is primary, as it describes the dynamics in terms of physical processes rather than purely mathematical constructs, such as the Lindblad operators are. This provides a new approach to quantum master equations, based on physical dynamics. 

We also note that our form of the master equation allows to uniquely define the free evolution Hamiltonian of the system. The same has been attempted previously through the ``minimum dissipation principle'' \cite{Hayden2022, Colla2022, Gatto2024}, and our work is in agreement with the approach, while providing a more robust physical reasoning for it. 

As our result was derived from the GLKS equation with no approximations, it holds true generally, including in the strong coupling regimes or far from equilibrium. This will provide a new pathway to the description of these typically theoretically complex cases. 

\subsection{The role of effective charge}
The central idea of the paper was the introduction of an effective charge, the physical quantity exchanged between the system and the bath, which does not necessarily commute with the Hamiltonian. 

The explicit inclusion of the exchange term makes our form of the quantum master equation perfectly suited for the description of systems exchanging multiple non-commuting quantities. Such effects, coined ``non-Abelian thermodynamics'' \cite{Guryanova2016, YungerHalpern2016, Lostaglio2017} have gained steady attention in the past decade, including experimental studies \cite{Kranzl2023}, however multiple open problems remain \cite{Majidy2023}. 

Our equation demonstrates that the exchange of non-commuting operators, or operators not commuting with the Hamiltonian is not purely the remit of rare and exotic effects, but instead is a much broader phenomenon. 

For a system with time-independent parameters exchanging energy with a bath, the only degree of freedom is coupling strength. Thus, the coupling strength is represented by the non-commutation of the effective charge, the ``exchange'' energy operator with the Hamiltonian, the free evolution energy operator. This is reflected in the experimentally observable -- the presence of a non-zero commutator between the ``exchange'' energy operator and the Hamiltonian implies the presence of uncertainty in the energy exchanged in every instance. This agrees with finite resonance widths being the key signature of strong coupling. If a system exchanges charge with the environment \cite{Site2024, Carlen2025, Reible2025}, strong coupling, or the non-commutation of $\N$ and $\Ham$ implies that the charge is not conserved in the system by its free evolution, which also agrees with the intuitive notion of strong coupling. 

We have to note a significant result in our work pertinent to the study of generalised charges. For the case of $\N$ not commuting with the Hamiltonian, the QME in our form (Eq.\ref{eq:exchange}) gives a stationary state with exponential terms in $\Ham$ and $\N$, but also the commutator(Eq.\ref{eq:case3-stst}). It can be written as:
\begin{equation}
\label{eq:gen-Gibbs}
    \dm_{st}=\frac{1}{Z}e^{-\beta \Ham + \mu \N +i\lambda [\Ham, \N]}
\end{equation}

This is in contradiction with the ``generalised Gibbs state'' assumed in non-Abelian thermodynamics \cite{Guryanova2016, YungerHalpern2016, Lostaglio2017}:
\begin{equation}
    \dm_{st}=\frac{1}{Z}e^{-\beta \Ham + \sum_i \mu_i \N_i}
\end{equation}

We believe our form is correct. The cited reasoning for the ``generalised Gibbs'' form above stems from the maximum entropy approach originated by Jaynes \cite{Jaynes1957, Jaynes1957-2, Mark2024}, where it is found through the maximisation of the system entropy under a number of constraints. However, we note that for two non-commuting operators, simultaneous constraints on their mean values are not possible. Therefore it is only reasonable for the thermal state to contain a term describing the quantum uncertainty between each two constraints as well.

\section{Summary and Conclusions}
\label{sec-conclusions}
The main result of the work is the introduction of a new framework for quantum master equations, which separates arbitrary dynamics given by the GKLS equation into the a sum of physical processes: free evolution, exchange of generalised charges, and pure dephasing. This reintroduces direct physical meaning into the purely mathematical structure of the GKLS equation. As the result was obtained from the GKLS equation structure directly, it is valid generally, with no assumptions on the system or the system-bath coupling.  
 
The concept of a generalised charge not necessarily commuting with the Hamiltonian unites the phenomena of strong coupling, particle exchange, and more general non-Abelian thermodynamics. The framework has shown several predictions supported by intuition and experiment, such as the finite linewidth for the strong coupling regime. Additionally, we have obtained a novel and significant result for the general form of the Gibbs distribution in the case of non-commuting generalised charges.    

These early results notwithstanding, the main aspect in the framework is currently lacking is physics. The result (Eq.\ref{eq:exchange}) was obtained by mathematical transformations of the GKLS equation, and therefore does not place any constraints of the rates $\gamma_p$ and $\gamma_m$. The most significant next step is the establishment of the extension of detailed balance \cite{Agarwal1973, Alicki1976, Scandi2026} beyond the regime for which it was established and into strong coupling conditions. From intuitive reasoning, the coupling strengths  $\gamma_p$ and $\gamma_m$, the uncertainty potential $\lambda$ in our form of the generalised Gibbs state (Eq. \ref{eq:gen-Gibbs}), and the degree of non-commutation between $\N$ and $\Ham$ should all be related.

One way to explore this crucial relation is to experimentally study the dynamics of a system in the strong coupling regime and explore the exceptional points, which allow to determine the system parameters \cite{am2015exceptional}. Such experiments have already been performed employing a superconducting qubit \cite{chen2022decoherence}.

Other directions include comparing the predictions of our theory with those of more standard approaches, such as the mean force Hamiltonian \cite{Anto-Sztrikacs2023, Burke2024} and setting the parameters for the two to agree. Should our desired relation be found, the theory will give a full approximation-free description of quantum dynamics beyond the standard theoretical limitations.

\section*{Acknowledgements}
E.P. is grateful to the Azrieli Foundation for the award of an Azrieli Fellowship.


\appendix
\section{GKLS equation transforms}
\label{app-transforms}
For ease of reference and the completeness of the work, here we list and prove all transforms that preserve the Lindblad equation form: 
\begin{equation}
    \frac{\dd \dm}{\dd t}=-i [ \Ham, \dm]+\sum \gamma_i \left(L_i \dm L_i^\dagger-\frac{1}{2}\{L_i^\dagger L_i, \dm\} \right)  
\end{equation}
\subsection{Energy shift}
The transformation: 
\begin{equation}
    \Ham \rightarrow \Ham - E \Id
\end{equation}
corresponds to the shift of the zero energy and does not affect the dynamics. 
\subsection{Lindblad operator scaling}
\label{ap-scaling}
\begin{equation}
    L_i \rightarrow \alpha_i L_i;\text{ } \gamma_i \rightarrow \frac{\gamma_i}{\alpha_i \alpha_i^*}
\end{equation}
The Lindblad jump operators can be scaled by a complex constant $\alpha$, with the rate coefficients corrected accordingly.  
\subsection{Lindblad operator -- identity shift}
The first non-trivial transform allows to shift terms between the jump operators and the Hamiltonian:
\begin{equation}
\label{eq:identity-theft}
    L_i \rightarrow L_i-\alpha_i \Id;\text{ } \Ham \rightarrow \Ham+\sum_i \frac{\gamma_i}{2i}\left(\alpha^* L_i-\alpha_i L_i^\dagger \right)
\end{equation}
\textbf{Proof:} We look at the transform for a single dissipative term in the sum (and thus drop the index $i$). Let $M=L-\alpha \Id$. Then the corresponding  dissipator term for $M$ is:
\begin{multline}
    \Miss(\dm)=M\dm M^\dagger -\frac{1}{2}\{M^\dagger M ; \dm \}=\\=(L-\alpha \Id)\dm (L^\dagger-\alpha^* \Id)-\frac{1}{2}\{(L^\dagger-\alpha^* \Id)(L-\alpha \Id); \dm \}
\end{multline}
After expanding the above, we arrive at:
\begin{multline}
    M\dm M^\dagger -\frac{1}{2}\{M^\dagger M ; \dm \}=\\=L\dm L^\dagger -\frac{1}{2}\{L^\dagger L ; \dm \}-\frac{1}{2} [\alpha^* L;\dm]+\frac{1}{2} [\alpha ^\dagger;\dm]
\end{multline}
which leads to Eq.\ref{eq:identity-theft}.
\subsection{Unitary transform}
Another non-trivial transform is unitary mixing of the Lindblad operators. 

First, applying the rescaling transform (Sec.\ref{ap-scaling}) we absorb the rate coefficients into the jump operators, to have the GKLS equation in the form:
\begin{equation}
    \frac{\dd \dm}{\dd t}=-i [ \Ham, \dm]+\sum \left(L_i \dm L_i^\dagger-\frac{1}{2}\{L_i^\dagger L_i, \dm\} \right)  
\end{equation}
Then, we show that the transform:
\begin{equation}
    M_i=\sum_k u_{ik}L_k
\end{equation}
where $U=u_{ik}$ is a unitary matrix, preserves the dissipator form.
\\[5pt]
\textbf{The ``sandwich'' terms:}
\begin{multline}
    \sum_i M_i \dm M_i^\dagger =\sum_i \left(\sum_k u_{ik}L_k\right)\dm \left(\sum_j u^*_{ij}L^\dagger_j \right)=\\=\sum_{kj}\left(\sum_i u_{ik}u^*_{ij} \right)L_k \dm L_j^\dagger=\\= \sum_{kj} \delta_{kj}L_k \dm L_j^\dagger=\sum_k L_k \dm L_k^\dagger
\end{multline}
since $U=u_{ik}$ is unitary, $U U^\dagger=\Id$ is equivalent to $\sum_i u_{ik}u^*_{ij}=\delta_{kj}$, where $\delta_{kj}$ is the Kroneker delta. 
\\[5pt]
\textbf{The anticommutator terms:}
\begin{multline}
    \sum_i M_i^\dagger M_i =\sum_i \left(\sum_j u^*_{ij}L^\dagger_j \right)\left(\sum_k u_{ik}L_k\right)=\\= \sum_{kj} \delta_{kj} L_j^\dagger L_k =\sum_k L_k^\dagger L_k 
\end{multline}
which, due to the linearity of the anticommutator gives:
\begin{equation}
    \sum_i \{ M_i^\dagger M_i; \dm\}=\sum_i \{ L_i^\dagger L_i; \dm\}
\end{equation}
therefore the ``sandwich'' and the ``anticommutator'' terms of the dissipator are preserved separately.
\section{Simple relations}
\label{ap:trivial}
Here, we prove some simple relations used in the mathematical derivation of our main result (Section \ref{sec-maths}).
\subsection{Properties of traceless operators}
The Pauli matrices, $\sigma_z$, $\sigma_x$, and $\sigma_y$, together with the identity matrix $\Id$, form a basis in the space of 2$\times$2 Hermitian operators, over the real numbers. If we restrict our considerations to traceless Hermitian operators, the linear space is three-dimensional, are the coefficients with $\sigma_z$, $\sigma_x$, and $\sigma_y$ are real. Moreover, any traceless 2$\times$2 operator $M$, not necessarily Hermitian, can be written as:
\begin{equation}
    M=m_z \sigma_z + m_x \sigma_x+ m_y \sigma_y
\end{equation}
where $m_z$, $m_x$, $m_y \in \mathbb{C}$. 

Then the following is true:
\begin{itemize}
    \item $M+M^\dagger \in \mathcal{H}$ and is Hermitian. 
    \item $M^2=(m_z \sigma_z + m_x \sigma_x+ m_y \sigma_y)^2=(m_z^2+m_x^2+m_y^2)\Id$, where $m_z^2+m_x^2+m_y^2$ is still, in general, complex-valued. 
    \item $\{M;M^\dagger \}=2(m_z m_z^*+m_x m_x^*+m_y m_y^*)\Id$ -- identity times a real number. 
    \\For two different operators, the commutator is still proportional to identity, but the coefficient is not necessarily real:
    \\
    $\{M;N \}=2(m_z n_z+m_x m_x+m_y n_y)\Id$
    \item   For the commutator:  
    \begin{multline*}
        [M;M^\dagger ]=2i ( (m_x m_y^*-m_x^* m_y)\sigma_z+\\+(m_y m_z^*-m_y^* m_z)\sigma_x+(m_z m_x^*-m_z^* m_x)\sigma_y )
    \end{multline*}
\end{itemize}
\subsection{Operator basis}
In the beginning of Section \ref{sec:equation}, we have elected a Hermitian orthonormal basis in the linear space of $L_1+L_1^\dagger$ and $L_2+L_2^\dagger$, $A_1$ and $A_2$, such that:
\begin{equation}
    \begin{split}
        A_1^2+A_2^2=\Id\\
        \Tr A_1 A_2=0
    \end{split}
\end{equation}
We need to complete it to the full three-dimensional space. 

To this end, we choose:
\begin{equation}
    A_3=\frac{1}{2i}[A_1; A_2]
\end{equation}
orthogonal to both $A_1$ and $A_2$. 

As $A_1$ and $A_2$ are traceless, $\{A_1; A_2\}\sim \Id$, and since $\Tr A_1 A_2=0$, the anticommutator is equal to zero: $\{A_1; A_2\}=0$. 

This gives: $A_1 A_2 =-A_2 A_1$, and, in turn:
\begin{equation}
    iA_3=i[A_1; A_2]=A_1 A_2=-A_2 A_1
\end{equation}
Finally, we check that $A_3$ is normalised:
\begin{multline}
    -4A_3^2=[A_1; A_2]^2=(2A_1 A_2) (-2 A_2 A_1)=\\=-4 A_1 A_2 A_2 A_1=-4 \Id
\end{multline}
which gives us: $A_3^2=\Id$.

\section{$\Ham\neq\N$, exceptional points}
\label{ap-exp}
The characteristic matrix of the master equation in the case of non-commuting $\Ham$ and $\N$ (the Hamiltonian of the system is equal to $E \N +\varepsilon \D$, see Eq.\ref{eq:case3}) is:
\begin{equation}
M_2=
    \begin{pmatrix}
    -\gamma & 0 & -\varepsilon\\[3pt]
    0 & -\frac{\gamma}{2} & E\\[3pt]
    \varepsilon & -E & -\frac{\gamma}{2}
\end{pmatrix}
\end{equation}
The eigenvalues, $\lambda_{1-3}$ satisfy the characteristic equation $\det(M_2-\lambda I)=0$, where $I$ is the identity matrix:
\begin{equation}
    \left(\gamma+\lambda \right)\left(\frac{\gamma}{2}+\lambda\right)^2+\varepsilon^2\left(\frac{\gamma}{2}+\lambda\right)+E^2 \left(\gamma+\lambda \right)=0
\end{equation}
Simplifying, and rescaling all energy parameters through $\gamma$ ($E\rightarrow E/\gamma$, $\varepsilon\rightarrow\varepsilon/\gamma$), we arrive at:
\begin{equation}
    \lambda^3 + 2\lambda^2+\lambda \left(\frac{5}{4}+\varepsilon^2+E^2 \right)+\left(\frac{1}{4}+\frac{\varepsilon^2}{2}+E^2 \right)=0
\end{equation}
As a cubic equation, it can always be analytically solved. If we define:
\begin{equation}
    \begin{split}
    &X=-\frac{1}{4}+3\varepsilon^2+3E^2\\
    &Y=-\frac{1}{4}+\frac{9\varepsilon^2}{2}-9E^2\\
    &Z=\sqrt{Y^2+4X^3}
\end{split}
\end{equation}
the general solution for the eigenvalues can be expressed as:
\begin{equation}
    \begin{split}
        &\lambda_1=-\frac{2}{3}-\frac{2^{\frac{2}{3}}X}{3\left(Y+Z \right)^{\frac{1}{3}}}+\frac{\left(Y+Z \right)^{\frac{1}{3}}}{3 \cdot 2^{\frac{1}{3}}}\\
        &\lambda_{2/3}=-\frac{2}{3}+\frac{(1\pm i\sqrt{3})X}{3\cdot 2^{\frac{2}{3}}\left(Y+Z \right)^{\frac{1}{3}}}-\frac{(1\mp i \sqrt{3})\left(Y+Z \right)^{\frac{1}{3}}}{6 \cdot 2^{\frac{1}{3}}}
    \end{split}
\end{equation}
For real $Z$ ($Y^2+4X^3>0$) all eigenvalues are real. If $Y^2+4X^3<0$, $\lambda_1$ is real, while $\lambda_2$ and $\lambda_3$ are complex and conjugates of each other. The transition $Y^2+4X^3=0$ corresponds to the second order exceptional points, in which $\lambda_2=\lambda_3$ and the eigenvectors coalesce. Finally, if $X=Y=0$ (in this case $Z=0$ as well), a triple degeneracy and a third-order exceptional point occurs. These manifest in cusps in the phase diagram \cite{Berry1979} (see Fig.\ref{fig-expoints}).

\providecommand{\noopsort}[1]{}\providecommand{\singleletter}[1]{#1}%

\end{document}